\documentclass[manuscript]{acmart}

\AtBeginDocument{%
  }

\setcopyright{acmlicensed}
\copyrightyear{2018}
\acmYear{2018}
\acmDOI{XXXXXXX.XXXXXXX}

\acmConference[Conference acronym 'XX]{Make sure to enter the correct
  conference title from your rights confirmation emai}{June 03--05,
  2018}{Woodstock, NY}
\acmISBN{978-1-4503-XXXX-X/18/06}

\usepackage{pifont}
\begin{document}

\title{DataScout: Automatic Data Fact Retrieval for Statement Augmentation with an LLM-Based Agent}
\author{Chuer Chen}
\affiliation{%
  \institution{Tongji University}
  \city{Shanghai}
  \country{China}
}

\author{Yuqi Liu}
\affiliation{%
  \institution{Tongji University}
  \city{Shanghai}
  \country{China}
}

\author{Danqing Shi}
\affiliation{%
  \institution{Aalto University}
  \city{Helsinki}
  \country{Finland}
}

\author{Shixiong Cao}
\affiliation{%
  \institution{Tongji University}
  \city{Shanghai}
  \country{China}
}

\author{Nan Cao}
\affiliation{%
  \institution{Tongji University}
  \city{Shanghai}
  \country{China}
}

\renewcommand{\shortauthors}{Trovato et al.}

\begin{abstract}
A data story typically integrates data facts from multiple perspectives and stances to construct a comprehensive and objective narrative. However, retrieving these facts demands time for data search and challenges the creator's analytical skills. In this work, we introduce DataScout, an interactive system that automatically performs reasoning and stance-based data facts retrieval to augment the user's statement. 
Particularly, DataScout leverages an LLM-based agent to construct a retrieval tree, enabling collaborative control of its expansion between users and the agent. 
The interface visualizes the retrieval tree as a mind map that eases users to intuitively steer the retrieval direction and effectively engage in reasoning and analysis. We evaluate the proposed system through case studies and in-depth expert interviews. Our evaluation demonstrates that DataScout can effectively retrieve multifaceted data facts from different stances, helping users verify their statements and enhance the credibility of their stories.

\end{abstract}




\begin{teaserfigure}
  \centering
  \includegraphics[width=\linewidth]{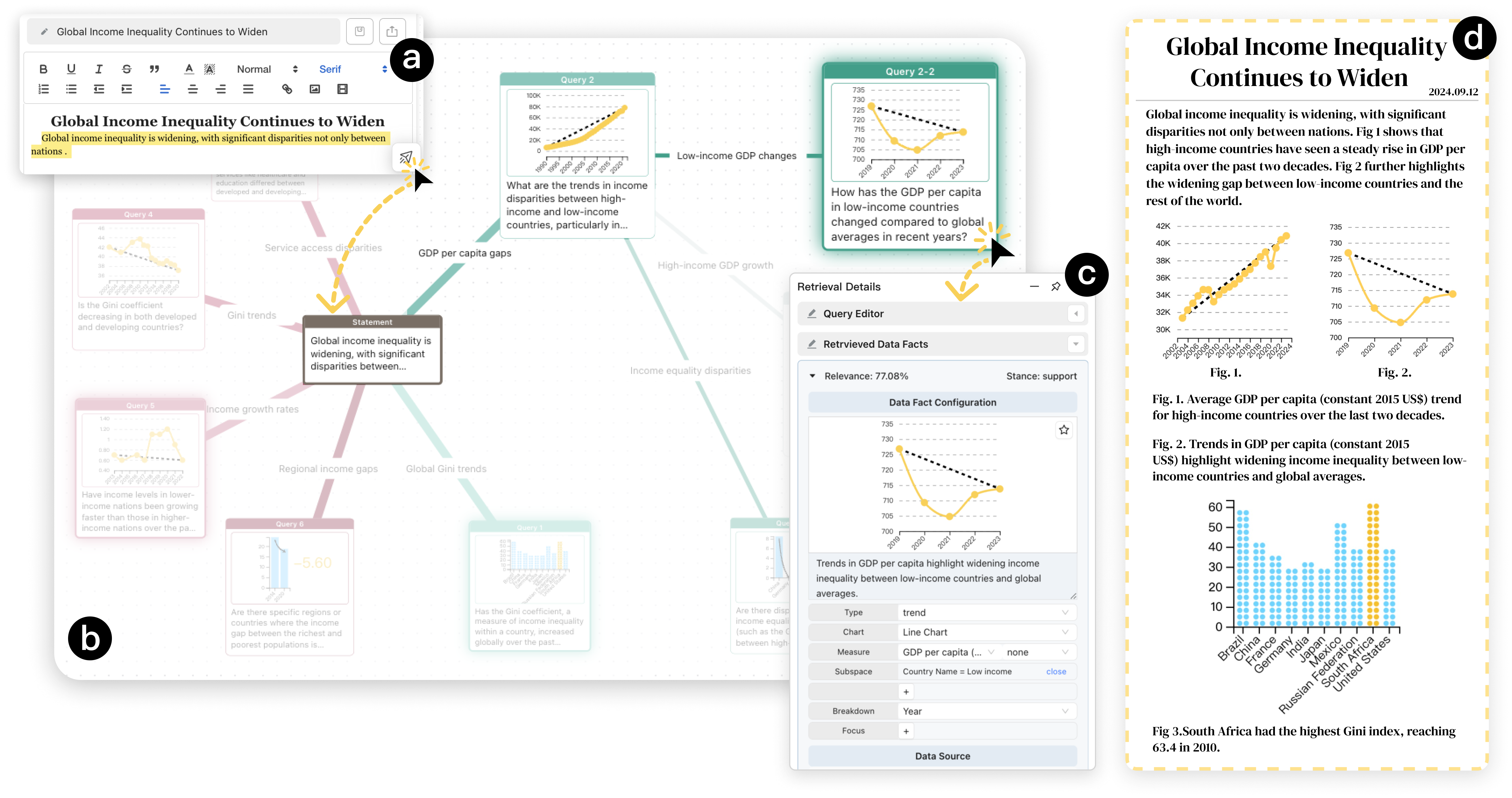}
  \caption{DataScout enables automatic stance-based data fact retrieval to enhance user statements. Generally, the retrieval process typically begins with (a) the user selecting a specific statement in the editor. The user can then (b) browse and expand the retrieval tree within the retrieval space view. Once a node is clicked, (c) the retrieval details view appears, showing the query and retrieved data facts, allowing the user to select appropriate facts and add them to the editor. An example data story about \textit{"Global income inequality continues to widen"} is shown in (d), with the three retrieved facts supporting the argument.} 
  \label{fig:teaser}
\end{teaserfigure}

\received{20 February 2007}
\received[revised]{12 March 2009}
\received[accepted]{5 June 2009}

\maketitle

\section{Introduction}
Data-driven storytelling, with its rich data content and expressive visualizations, has become a widely adopted method of information dissemination in the digital era. It often goes beyond merely presenting data to focus on persuading the audience through effective argumentation, which involves presenting, supporting, reinforcing, contradicting, or discussing a given statement~\cite{DataDrivenStorytelling}. To achieve compelling argumentation, it is crucial to present accurate and relevant data facts~\cite{datafacts2019} in the story to augment the statement thereby enhancing the credibility of the narrative.

However, retrieving comprehensive data facts is a complicated and labor-intensive process that demands expertise in data analysis. To alleviate creator's burdens, researchers in data visualization have explored various methods for automatic data facts generation~\cite{datashot,shi2020calliope,Lu2021AutomaticGO}. For example, DataShot~\cite{datashot} extracts data facts from the tabular data by computing various facts within each enumerated subspace and scoring them for ranking. The enumeration method generates many redundant data facts, and manual scoring rules struggle to adapt to complex semantic environments. Recent advances in Large Language Models (LLMs) offer new opportunities for automating data analysis. Some studies~\cite{shen2024datastoryautomaticanimated, cheng2023gpt} employ LLMs with carefully crafted prompts to extract insights from data. Although the strong understanding and reasoning abilities of LLMs enable efficient fact extraction, they still cannot extract facts that align with a specific argumentative direction. On the other hand, most studies are limited to extracting facts from the dataset provided by the user and cannot actively retrieve multidimensional data to strengthen the argument. Users must invest significant effort in data search and then apply data fact extraction methods to complete the entire workflow. While recent works have enabled fast task-driven search~\cite{Wang_Fernandez_2023, Herzig_Müller_Krichene_Eisenschlos_2021}, there remains a need for an iterative interface~\cite{datasetsearch2024} that allows users to dynamically adjust the search direction based on argumentative needs.

To better understand the workflow and challenges of retrieving data facts, we conducted an interview-based formative study with four data storytelling experts. The formative study found that retrieving data facts is a challenging and time-consuming task. To enhance the credibility of the story, creators generally need to prepare comprehensive data facts. This requires breaking down search objectives into keywords that allow for the retrieval of diverse data sources, and then analyzing the collected data to extract valuable facts that will strengthen the argument.

To address the above challenges, we introduce DataScout, an intelligent system designed to automatically retrieve data facts for statement augmentation. In particular, we propose an LLM-based agent that collaborates with users to jointly construct a retrieval tree based on argumentative needs. The retrieval tree takes the statement to be augmented as the root node and expands nodes based on the user’s expected stance, with each node containing a query and its retrieved facts. Users simply select the node and desired stance, and the agent executes the stance-based data fact retrieval automatically: First, the query decomposition module breaks down the selected node's query into multifaceted sub-queries. Then, the data search module performs text-to-SQL generation and filters sub-tables related to sub-queries. Finally, the fact extraction module uses Chain-of-Thought (CoT) reasoning to extract facts from sub-tables, and further evaluates the relevance and predicted stance of them to assist users in assessing the facts. Additionally, the planning module of the agent will recommend a node worth further exploration based on the retrieval results, guiding users in choosing the next argumentative direction. All the modules are instructed to consider the expected stance through carefully crafted prompts to ensure that the retrieved facts align with the user's argumentative needs. The system also features an iterative interface with a mind map layout retrieval space, which allows users to clearly view the retrieval results and efficiently manage the tree expansion.

In summary, we make the following contributions:
\begin{itemize}
  \item We present DataScout, a system that enables stance-based data facts retrieval for augmenting the user's statement with an LLM-based agent. The system allows users to write a statement and retrieve supporting or opposing facts within a mind map layout space.
  \item We conduct a formative study with four data storytelling experts that identifies the workflow and challenges of retrieving data facts. 
  \item We demonstrate the utility of the proposed system via interviews and case studies with three expert users.
\end{itemize}

\section{Related Work}
In this section, we discuss prior studies that are closely related to our work, namely, data-driven storytelling, information retrieval system, and natural language interfaces for visualization.

\subsection{Data-driven Storytelling}
Data-driven storytelling is a technique that constructs narratives supported or enhanced by data, often utilizing visualizations to vividly integrate data into the story's presentation~\cite{riche2018data}. This method has gained prominence in visualization, driven by the need to effectively communicate complex data and leading to the development of advanced techniques~\cite{tong2018storytelling}. These techniques aim to create structured visual narratives that facilitate data comprehension and improve memory retention~\cite{recall2016beyond,hullman2013deeper,dove2012narrative,segel2010narrative}.

Theoretical design spaces of data-driven storytelling have been explored in previous studies~\cite{segel2010narrative,hullman2013deeper,mckenna2017visual,stolper2016emerging,bach2018narrative}, focusing on the integration of data into narratives. Segel and Heer~\cite{segel2010narrative} constructed a design space and identified seven genres of narrative visualization. Hullman et al.~\cite{hullman2013deeper} outlined key design actions in the storytelling process. McKenna et al.~\cite{mckenna2017visual} identified seven 'flow-factors' that contribute to visual narrative flow. Stolper et al.~\cite{stolper2016emerging} categorized techniques into four main areas to enhance storytelling. Bach et al.~\cite{bach2018narrative} analyzed 18 narrative patterns and demonstrated how they aid storytellers in planning and presenting their stories.

Guided by these theoretical insights, the development of effective authoring tools can be significantly advanced. Some tools enable interactive creation and positioning of custom annotations to build visual narratives~\cite{ren2017chartaccent,wang2018infonice,bryan2016temporal,conlen2021idyll}. Other tools generate valuable data narratives by organizing meaningful sequences or layouts of visualizations~\cite{zhao2021chartstory,conlen2021idyll,shi2019task}. To improve efficiency, certain tools automate the generation of data stories~\cite{shi2020calliope,shi2021autoclips,sun2022erato,chen2018supporting} by identifying data facts from the datasets and organizing them into q cohesive narrative. 

However, most existing tools generally assume that the data is already available, overlooking the process of searching for relevant data. In contrast, DataScout can automatically search for multifaceted data to augment users' statements and extract data facts from the data for visualization, which further simplifies the authoring of data stories.

\subsection{Information Retrieval System}
Traditional information retrieval (IR) systems have primarily focused on efficiently locating relevant information through techniques like keyword matching~\cite{siddiqui2005integrating,uthayan2015hybrid}, ranking algorithms~\cite{robertson2009probabilistic,jain2011page,szummer2011semi}, and query processing~\cite{jain2021fuzzy,kaur2021query,bai2007using}. However, the rise of large language models (LLMs) is shifting the paradigm from searching to generating information~\cite{shi2023replug,zhu2023large,borgeaud2022improving}, revealing the limitations of traditional IR systems in terms of interactivity and architecture~\cite{tang2024self}.

This paradigm shift has inspired a range of approaches that leverage advanced retrieval techniques and pipelines to enhance interactivity and search results. For example, 
DataDive~\cite{kim2024datadive} employs an LLM-driven pipeline to generate questions from statistical reports and retrieve datasets, though it lacks a non-linear interface. Self-Retrieval~\cite{tang2024self} is an LLM-driven information retrieval architecture that integrates necessary information retrieval (IR) capabilities within a single LLM, leveraging its full potential throughout the IR process.


Unlike the systems mentioned above, DataScout visualizes the retrieval process in a mind-map layout within the user interface, allowing users to clearly follow the reasoning behind it and work with the LLM-based agent to collaboratively steer the retrieval direction.

\subsection{Natural Language Interfaces for Visualization}

With advancements in Natural Language Processing (NLP) technology ~\cite{belinkov2019analysis,young2018recent}, there has been a notable increase in the development of Natural Language Interfaces (NLI) for Visualization ~\cite{shen2022towards}. Natural Language Interfaces offers an intuitive method for exploratory visual analysis, allowing users to pose questions about input data in natural language and receive corresponding data-related answers in visual formats ~\cite{gao2015datatone,narechania2020nl4dv,setlur2016eviza}.

A common challenge for NLIs is how users' intentions expressed in natural language are accurately interpreted ~\cite{Guo2024Talk2Data}. To improve the accuracy of natural language interpretation, existing NLIs are designed either to help users formulate more precise natural language queries ~\cite{cox2001multi,setlur2016eviza,yu2019flowsense} or to resolve ambiguities in the input queries ~\cite{gao2015datatone,sun2010articulate,wen2005optimization,narechania2020nl4dv}. This enhances the system's ability to understand user requirements and drive more accurate data analysis for answering questions. However, these NLIs are likely to encounter parsing failures when addressing complex issues ~\cite{yu2019flowsense}.

Recently, large language models (LLMs) have demonstrated remarkably advanced reasoning capabilities through in-context learning ~\cite{brown2020language,dinh2022lift}. Leveraging this strong performance in a wide range of natural language tasks ~\cite{qin2023chatgpt,gilardi2023chatgpt,sun2023chatgpt} and complex data analysis tasks ~\cite{cheng2023gpt,savelka2023can}, recent studies, such as Chat2VIS ~\cite{maddigan2023chat2vis}, LIDA ~\cite{dibia2023lida}, and ChartGPT ~\cite{tian2024chartgpt}, investigate the innovative use of LLMs to enhance NLIs for data visualization.

Inspired by these LLM-powered studies, DataScout employs LLMs to retrieve data facts based on expected stance, which takes a statement and a stance as natural language input and outputs relevant data facts along with corresponding visualizations to support or oppose the given statement.

\section{Formative Study}
We conducted an interview-based study to understand current practices and limitations of how creators discover data facts. Based on the findings, we propose design goals to help creators efficiently extract relevant data facts.

\subsection{Procedure}
Semi-structured interviews were conducted with four domain experts (E1-E4, 3 females, 1 male), including a visualization researcher, a journalism editor, and two data story creators, each with an average of 5.75 years of experience in data storytelling. 

The interviews began with questions about the participants' backgrounds and their experience in authoring data stories. Participants were then asked to detail their workflows for retrieving data facts during the authoring process and to identify any challenges they faced. Each interview lasted approximately one hour and was conducted online via meeting software. After completing the interviews, we summarized the findings about the workflow and common challenges. 

\subsection{Findings}
\textbf{\textit{Workflow for Retrieving Data Facts.}} The general workflow employed by the participants consists of three main steps: First, given the topic of the data story, creators learn context information of the topic and then outline a content framework for the story. In the second step, they search the internet using keywords based on the content framework and retrieve relevant datasets from sources such as official government websites, third-party dataset platforms (e.g., Kaggle), news, academic papers, and web scraping. Finally, they clean and filter the collected data, then use data analysis tools (e.g., Python, Excel) to uncover valuable data facts from the data.

\textbf{\textit{Data facts should be comprehensive.}} When discussing the factors to consider when retrieving data facts, participants emphasized the importance of covering multiple perspectives to ensure the comprehensiveness of the data story. For instance, E1 mentioned the need to search for data as thoroughly as possible. On the other hand, while data facts need to align with the topic, it is crucial to maintain objectivity. As E4 noted, \textit{"The topic of a data story might imply a certain stance, and some data facts can support that topic. However, creators should incorporate data facts that represent all stances reflected in the original data to ensure the objectivity of the story."}

\textbf{\textit{Extracting Data Facts is Challenging.}} Participants reported that extracting meaningful data facts from large datasets is challenging, especially for novices with limited experience. As E1 noted, \textit{"Some of the novices I've worked with lack basic knowledge about data facts. For example, when analyzing trends, they don't realize that one column must be temporal data."} In addition to novices, experienced creators are challenged by how to reason out the next data fact. Data facts in a storyline need to come from multiple perspectives and be logically connected. E3 noted that after identifying data fact A, she sometimes struggles to find additional data facts that are logically connected to it. 

\textbf{\textit{The Workflow is Time-Consuming.}} Participants stated that retrieving data facts from large datasets is a time-consuming and tedious process. Data search usually consumes $10\%-30\%$ of the data storytelling process, requiring creators to try various search strategies before finding the appropriate datasets. E3 shared her experience searching the proportion of female athletes: \textit{"I tried using the event name and year as keywords, and also checked the official websites of each edition, taking many detours before finally finding the data."} Additionally, data cleaning and analysis are equally laborious, with creators spending around $30\%$ of their time using tools (e.g., Excel, Tableau) and programming languages (e.g., Python) for data analysis. 
 
\subsection{Design Goals}
Based on the findings, we established three goals to guide the design of DataScout to help creators efficiently reason and retrieve data facts relevant to the topics.

\begin{enumerate}
\item[{\bf D1}]{\bf Supporting Stance-Based Retrieval.} Data facts from various viewpoints can enhance the credibility of a data story. However, creators' subjective impressions may introduce biases in retrieved data facts. Therefore, DataScout aims to support stance-based retrieval to minimize potential bias and assist users in exploring data comprehensively, thereby creating a balanced and nuanced narrative.

\item[{\bf D2}] {\bf Deriving Search Strategies.} Searching for suitable datasets within extensive data requires advanced search skills and reasoning abilities. To improve search efficiency, DataScout should intelligently devise optimal search strategies by generating diverse search queries based on user-input statements. 

\item[{\bf D3}] {\bf Automatically Extracting Data Facts.} DataScout should automatically extract meaningful data facts to lower the barrier for novices. This feature should not only extract valid and reasonable data facts but also ensure alignment with the given statements and stances. Additionally, it's necessary to provide evaluations of the data facts to enhance the explainability of the extraction results, thereby helping users assess the reliability of the data facts. 

\end{enumerate}

\section{DataScout}
To fulfill the design goals, we present DataScout, a system that supports retrieving data facts based on stances and statements by leveraging an LLM-based agent powered by GPT-4o~\cite{openai_gpt4}. In this section, we first introduce the design of the system. After that, we present the details of the proposed agent.

\subsection{System Overview}
The design of the DataScout system consists of two modules: (1) the LLM-based agent and (2) the user interface. The system organizes the retrieval results into a retrieval tree, displayed in a mind-map layout retrieval space within the user interface, which visualizes the reasoning process behind the retrieval. The tree's expansion is collaboratively controlled by the user and the LLM-based agent. When the user inputs a statement, the agent automatically retrieves data facts that support and oppose the statement (\textbf{D1}) as the first-level nodes of the retrieval tree for further exploration. When the user selects a node, the agent first decomposes the query into multiple sub-queries to search for multifaceted data (\textbf{D2}). Next, the agent executes the data search module to find relevant sub-tables for each sub-query. Then, using Chain-of-Thought (CoT) prompting, the fact extraction module extracts stance-aligned data facts from the sub-tables (\textbf{D3}), evaluates them for relevance and stance, and ranks them. All sub-queries and retrieved facts are incorporated into the retrieval tree as child nodes. Finally, the planning module recommends one of the newly generated nodes for further expansion, guiding the user’s next step in the retrieval process.


\subsection{Problem Formulation}
The task of the agent is to augment the statement by constructing a retrieval tree that retrieves data facts aligned with the user-input statement and stances. The root of the tree represents the input statement, while each child node includes a generated sub-query and corresponding data facts. We define the expansion of this tree as a Partially Observable Markov Decision Process (POMDP)~\cite{spaan2012partially, hazra2024saycanpay}, denoted as a tuple: 


\begin{equation}
    (\mathcal{S},\mathcal{O}, \mathcal{A}, \mathcal{P}, \mathcal{G}, \mathcal{R})
\end{equation}

where $\mathcal{S}$ is the state space; $\mathcal{O}$ is a set of observations retrieved from states; $\mathcal{A}$ is the set of
actions; $\mathcal{P}$ is the transition model; $\mathcal{G}$ is the set of goal states; $\mathcal{R}$ is the reward function.

In our settings, the state space $\mathcal{S}$ refers to the entire retrieval tree, while actions $\mathcal{A}$ are defined as the user's selection of a node to expand. Observations $\mathcal{O}$ consist of the newly generated nodes following each expansion. The transition model $\mathcal{P}$ describes the probabilities of transitioning between states as the user and the agent interacts with the retrieval tree. $\mathcal{G}$ represents the specific objectives related to exploratory value for the retrieval task. The reward function $\mathcal{R}$ is quantified by the number of retrieved data facts that are highly relevant and align with the expected stance.

Initially, we construct a database comprising large datasets from the World Bank's World Development Indicators and consider it, along with the retrieval tree and user interactions, as the environment $E$.

\begin{figure}[!tbh]
  \centering
  \includegraphics[width=\linewidth]{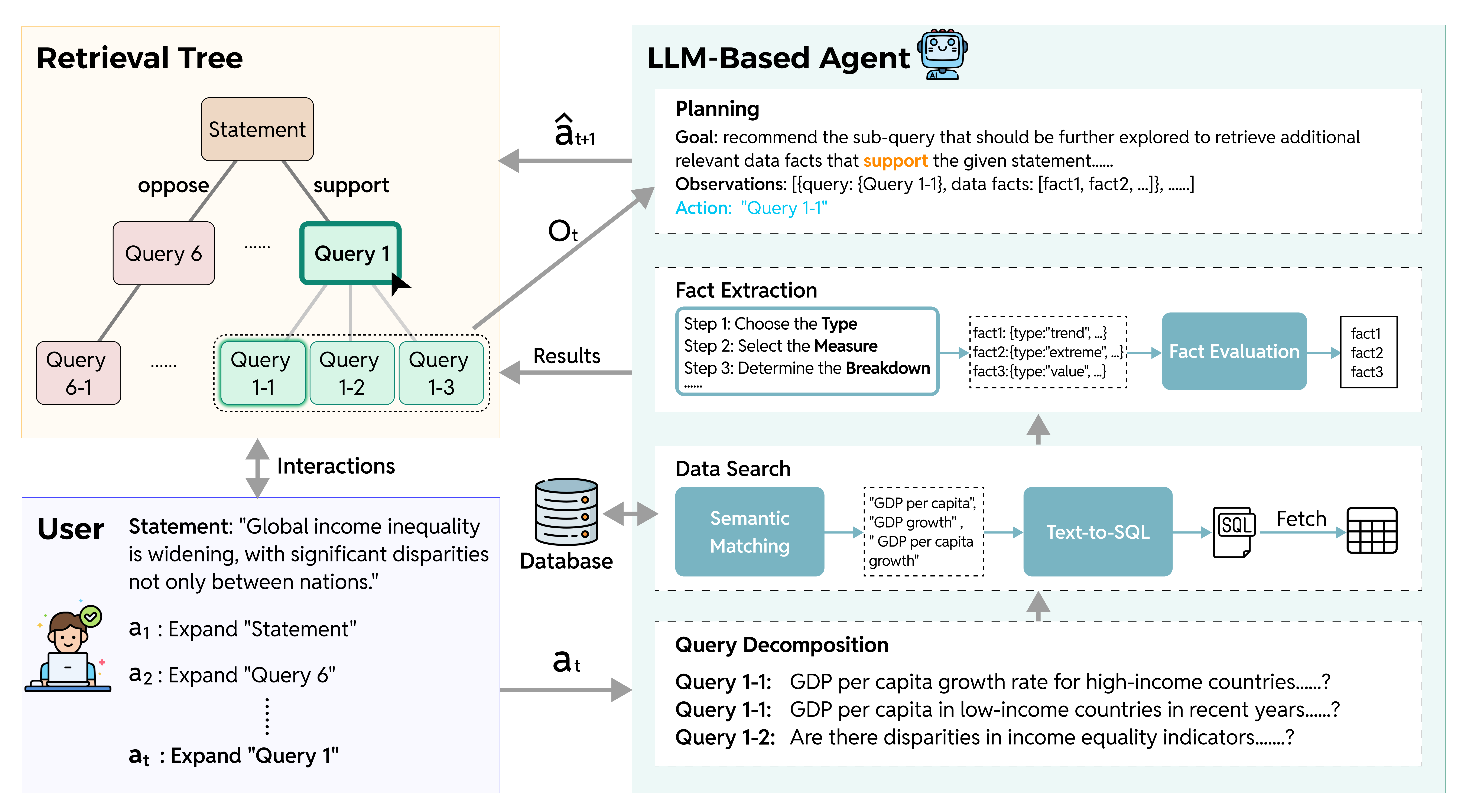}
  \caption{An overview of the LLM-Based agent. When the agent receives the user's instruction to expand a node, it sequentially executes the query decomposition, data search, and fact extraction processes. The generated sub-queries and retrieved data facts are then returned to the retrieval tree as new child nodes. Based on these child nodes, the agent plans the next node worth expanding to assist the user in decision-making.} 
  \label{fig:pipeline}
  \vspace{-1.5em}
\end{figure}

\subsection{Agent Design}
The LLM-based agent consists of four modules (Fig.~\ref{fig:pipeline}), including the planning module for recommending the retrieval direction, the query decomposition module for generating sub-queries, the data search module for finding relevant datasets, and the fact extraction module for uncovering and evaluating data facts. The planning module and the user collaboratively make the decision on the optimal action, which is then executed in sequence by the remaining three modules.

\subsubsection{Planning}
The planning module collaborates with the user to control the expansion of the retrieval tree. At each time step $t$, the user selects a node $n_i$ and inputs the desired stance ${stance}_t$ as the action $a_t \in A$ for expansion. The agent takes these inputs and performs query decomposition, data search, and fact extraction to complete the retrieval process, resulting in $m$ child nodes $\{n_{i,1}...n_{i,m}\}$, each of which contains a sub-query and the data facts retrieved to demonstrate $stance_t$. The agent then receives these child nodes, the desired stance, and the statement as the observation $O_t=\{n_{i,1}...n_{i,m}, stance_t, \text{statement}\}$. We set the goal $g_t \in \mathcal{G}$ as recommending the child node with the highest potential for retrieving high-quality data facts to guide the LLM's reasoning and planning process. 
The LLM, denoted as $\pi$, then generates a recommended child node as an intermediate decision:
\begin{equation}
    \hat{a}_{t+1} = \pi(O_t, g_t)
\end{equation}

The user will refer to the recommended child node to take the final action $a_{t+1}$. This human-AI collaborative approach integrates the LLM's reasoning and planning capabilities with human intelligence to achieve efficient and effective expansion of the retrieval tree.  

\subsubsection{Query Decomposition}
The query decomposition module leverages a few-shot prompting technique to guide the LLM in converting the query of the current node into detailed sub-queries from various perspectives, which facilitates data search by encouraging the agent to explore diverse angles. Unlike existing query decomposition methods, this module generates sub-queries based on the input stance and statement, making it easier to retrieve data facts that support the stance and augment the statement. As shown in Fig x, we provide two examples in the prompt that demonstrate how to generate sub-queries for the same statement under two different stances. With these examples, we can steer the LLM to better understand how to generate sub-queries that align with the stance and statement, enhancing the accuracy of data fact retrieval. 


\subsubsection{Data Search} The strategy of the data search module is to find relevant datasets in the database for each generated sub-query. This process is carried out in three steps: First, we use a sentence-transformers model~\cite{reimers-2019-sentence-bert}, all-MiniLM-L6-v2, to map the sub-query and all data fields in the database into embeddings. Then, we calculate and rank the semantic similarity between the query embedding and all pre-computed data field embeddings, selecting the top three most similar data fields and extracting the datasets they belong to as candidate datasets. In the second step, we apply an LLM-based text-to-SQL~\cite{rajkumar2022evaluatingtexttosqlcapabilitieslarge} method to generate SQL queries for each candidate dataset in order to filter out the sub-tables most relevant to the query. Finally, we execute the generated SQL queries and refine them if any errors occur during execution. The resulting sub-tables are used for subsequent data fact extraction.

\subsubsection{Fact Extraction} A data fact is a fundamental building block of the data story, which represents a particular type of numerical property on a subspace of the dataset. We follow the definition introduced in~\cite{shi2020calliope} to formalize the data fact into a 5-tuple structure:
\begin{equation}
    f=\{type, subspace, breakdown, measure, focus\}
\end{equation}

where $type$ indicates the type of information described by the fact, including \textit{value}, \textit{difference}, \textit{proportion}, \textit{trend}, \textit{categorization}, \textit{distribution}, \textit{rank}, \textit{association}, \textit{extreme} and \textit{outlier}; $subspace$ is a set of data filters that restricts the data scope of the fact; $breakdown$ is a temporal or categorical data field that divide the subspace into groups; $measure$ is a numerical data field that can be aggregated by \textit{count}, \textit{sum}, \textit{average}, \textit{minimum}, or \textit{maximum} methods within the subspace or each data group; $focus$ indicates one or more groups to highlight.

Previous works on LLM-based visualization have demonstrated superior performance in generating visualizations~\cite{maddigan2023chat2vis, tian2024chartgpt, dibia2023lida, wu2024NLLLM} and extracting data insights~\cite{cheng2023gpt, shen2024datastoryautomaticanimated, ma-etal-2023-insightpilot}. Therefore, the module leverages LLM's advanced reasoning ability to extract data facts from the sub-tables that align with the stance to enhance the given statement. Inspired by ~\cite{Guo2024Talk2Data}, we use Chain-of-Thought (CoT)~\cite{wei2023chainofthoughtpromptingelicitsreasoning} prompting technique to progressively select each item of the fact. 

CoT performs a step-by-step reasoning process that enables LLMs to perform complex reasoning. Extracting data facts is a challenging data analysis task due to the interrelated nature of each item. For instance, the choice of type can affect the permissible values for other items; if the \textit{type} is set to \textit{"trend"}, the \textit{breakdown} must be a temporal data field. Hence, using CoT helps the LLM better understand the structure of the fact and extract valid facts. Specifically, the designed prompt directs the LLM to make selections in the sequence of \textit{type}, \textit{measure}, \textit{breakdown}, \textit{subspace}, and \textit{focus}, with each step clarifying the definition and rules of the item. In addition to generating facts, the prompt also instructs the LLM to generate descriptions for each fact. To extract diverse facts, the prompt guides the LLM to repeat this reasoning process until three facts are generated for each sub-table. 

After the extraction process, each node gets a list of generated facts. To help users better assess the usefulness of each fact, we evaluate facts by introducing relevance and stance probability, and rank them according to these indicators. Particularly, relevance reflects how the meaning of a fact relates to the statement and query. We calculate the semantic similarity between the fact description embedding and both the statement and query embeddings, and take the average as the relevance score. For stance probability, we apply the LLM to predict the stance of the fact towards the statement as the label, generating both a label and a probability value. Then, we group facts by predicted stance label, prioritize the group matching the input stance, and rank facts within each group by relevance, which ensures that the most relevant and stance-aligned facts are highlighted for users. 

If no sub-tables are provided or valid facts cannot be extracted, the fact list will be empty. Additionally, although we instruct the LLM to consider stance during fact extraction, it's possible that all facts reflect the contrary stance. In such cases, the node and facts will still be retained in the retrieval tree to ensure the integrity and objectivity of the retrieval.

\section{User Interface}
\begin{figure}[!tbh]
  \centering
  \includegraphics[width=\linewidth]{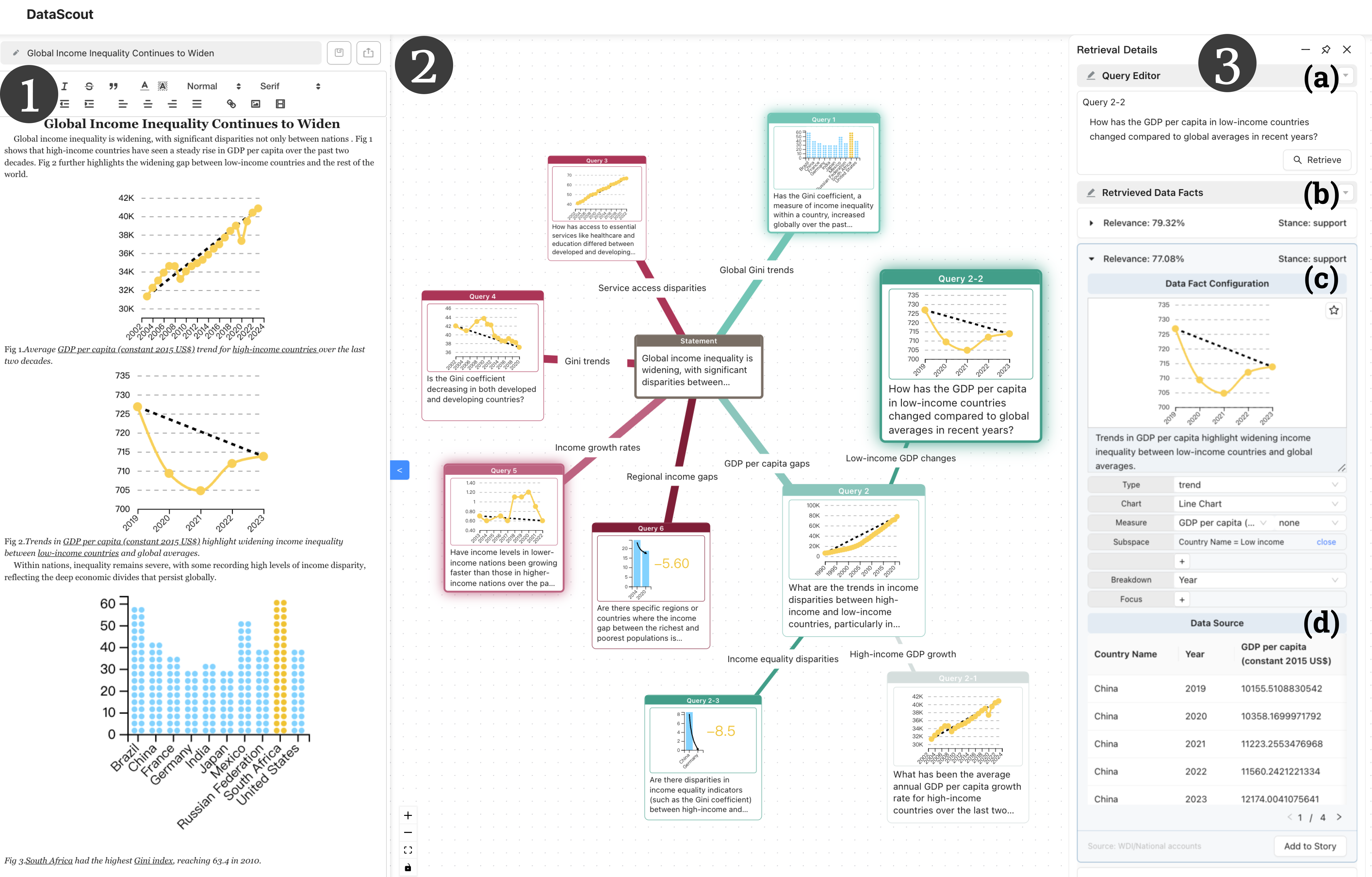}
  \caption{The interface consists of three major views: {\fontsize{10pt}{10pt}\selectfont \ding{182}}  the editor view; {\fontsize{10pt}{10pt}\selectfont \ding{183}} the retrieval space view; {\fontsize{10pt}{10pt}\selectfont \ding{184}} the retrieval details view.} 
  \label{fig:interace}
  \vspace{-1.5em}
\end{figure}

As shown in Fig.~\ref{fig:interace}, the interface of DataScout consists of three major views: (1) the editor view (Fig.~\ref{fig:interace}-1) for writing data stories and selecting a statement for retrieval; (2) the retrieval space view (Fig.~\ref{fig:interace}-2) displaying the retrieval tree as a mind map and enabling collaborative expansion; and (3) the retrieval details view (Fig.~\ref{fig:interace}-3) presenting the query and retrieved data facts for a selected node. In this section, we introduce the details of the views and present a usage scenario to illustrate how a user retrieves data facts in the system.


\subsection{Editor}
DataScout provides an editor (Fig.~\ref{fig:interace}-1) that allows users to create, export, and save data stories using rich text editing features. Additionally, when users select a segment of text, a retrieval button appears. Clicking this button initiates the retrieval process. Once users identify satisfactory data facts in the retrieval space and retrieval detail interface, they can choose to add these facts to the editor, aiding in the creation of expressive data stories.

\subsection{Retrieval Space}
DataScout provides an interactive retrieval space (Fig.~\ref{fig:interace}-2) that presents the retrieval tree in a mind map layout, which facilitates users to clearly view the retrieval results and efficiently control the tree's expansion. A mind map is a diagram that visually organizes information into a hierarchy~\cite{hopper2007mapping}, making it suitable for complex and multifaceted information retrieval tasks. In the retrieval space, the root node represents the statement selected by the user in the editor for augmentation. Each child node displays the query and the visualization as a card, with the default view showing the visualization of the top-ranked fact in the fact list. To make the structure clearer, nodes supporting the statement are placed on the right side of the root node, while nodes opposing the statement are placed on the left. Users can click the plus button on a node card to expand and retrieve facts based on the stance of that node. 

In addition to the layout design, we use two types of visual encoding to help users quickly assess the value of a node: (1) \textbf{Color encoding} maps the stance to the node's color (green for support, red for oppose) to clearly distinguish between stances, with higher stance probability resulting in more saturated color. (2) \textbf{Size encoding} maps relevance to the node's size, with higher relevance score leading to a larger node. Here, the relevance and stance probability of the top-ranked fact is considered the node's result and applied for encoding. 

We also design the following interactions to improve user experience and facilitate efficient retrieval: (1) highlighting the border of agent-planned node to guide users in selecting the next action; (2) displaying the query keywords on the edge, enabling users to modify the keywords and retrieve the node again; and (3) generating an unobstructed initial layout automatically, while allowing nodes to be dragged, collapsed, or expanded to adjust the layout.

\subsection{Retrieval Details}
The Retrieval Details interface is accessed by double-clicking a specific node in the retrieval space. This interface supports users in modifying queries, viewing, selecting, and editing data facts. It is divided into two main sections: Query Editor ((Fig.~\ref{fig:interace}-(a)) and Retrieved Data Facts (Fig.~\ref{fig:interace}-(b)).

In the Query Editor, users can view the automatically generated query. If needed, they can edit the query and click the "Retrieve" button to perform the data search and fact extraction again based on the new query.

The Retrieved Data Facts panel displays a list of data facts retrieved by the query, sorted by the relevance score and predicted stance. Each data fact is shown with two components: Data Fact Configuration (Fig.~\ref{fig:interace}-(c)) and Data Source (Fig.~\ref{fig:interace}-(d)). Data Fact Configuration presents the structural information and description of the data fact, which users can edit. Data Source shows the sub-table from which the fact is extracted, enabling users to verify the origin of the data, assess its reliability, and have confidence that the facts are grounded in authentic and credible sources.

\subsection{Usage Scenario}
Zoe, a journalist with profound insights into global economic trends, noticed a trending topic online: "\textit{Global income inequality is widening, with significant disparities between nations.}" Zoe decided to delve deeper into this phenomenon, aiming to create a data story on global income inequality through data visualization and narrative techniques, highlighting the gravity of this global issue to the public. However, without a deep understanding of specific data related to this global issue, creating this data story poses significant challenges.

Introduced with the DataScout system, Zoe writes this statement, "\textit{Global income inequality is widening, with significant disparities between nations.}" in the editor (Fig.~\ref{fig:interace}-1) on the left side of the DataScout system. After selecting this sentence, she clicks the retrieval button. Recognizing Zoe's interest in this topic, the DataScout system generates six child nodes branching out from the central topic based on the stance of supporting or opposing the statement in the retrieval space (Fig.~\ref{fig:interace}-2).  Each node contains a query, data from authoritative databases, thought-provoking data facts, and corresponding descriptions. The retrieval space uses node size to represent the relevance of the data fact and color intensity to indicate the degree of support or opposition. Moreover, it highlights the nodes worth expanding further. 


Zoe first browses and observes the six child nodes. She finds that from a supportive standpoint, both the Gini coefficient and GDP per capita are compelling angles to illustrate this topic. She double-clicks one of the largest nodes (\textit{Query 2}), which shows the query: \textit{"What are the trends in income disparities between high-income and low-income countries, particularly in terms of GDP per capita?"} Then, the corresponding retrieval details (Fig.~\ref{fig:interace}-3) view pops up, with the facts describing GDP per capita trends across various countries from 1990 to 2021. She feels that these data facts provide limited support for the statement and desires more in-depth information.


Zoe further expands this node by clicking the add button on the card, leading to results on "High-income GDP growth", "Low-income GDP changes" and "Income equality disparities". The first two results capture her attention, as they reveal the recent years' changes in GDP per capita (constant 2015 US\$) in high-income and low-income countries, which robustly support the statement. Through comparisons of multiple retrieved data facts (Fig.~\ref{fig:interace}-3(b)), Zoe adds these two data facts to the editor as key elements of her data story.

Zoe continues to check another node (\textit{Query 1}) that supports the statement, with the first fact showing that South Africa had the highest Gini index among the ten countries. She believes this extreme-type fact provides a strong contrast to highlight global inequality. Therefore, she also adds this fact to her data story. She uses the editor's formatting tools to organize the content and finally generates a compelling data story.

In the whole workflow, Zoe also observes the data facts under the opposing stance. Although she chooses not to delve deeper, she acknowledges that they provided her with some reflection and insights. For instance, the rising trend in educational attainment in some developing countries is a noteworthy aspect that she believes deserves attention.

\section{Evaluation}

To verify the overall utility of DataScout, we conducted case studies together with semi-structured interviews with three experts in the domain of data storytelling. Two experts (E3, E4) had participated in the formative study. Both are senior creators, with E3 having 5 years of experience in data storytelling and E4 having 3 years. The third expert (E5) is a fifth-year PhD candidate specializing in natural language-driven data visualization.

\subsection{Database Preparation}
In this evaluation, we constructed a database using the World Development Indicators dataset~\cite{worldbank_wdi} compiled by the World Bank from officially recognized international sources. This database includes 1,492 series covering education, economy, health, environment, and other aspects across countries, making it highly suitable for assessing our data fact retrieval system. Additionally, we designed three statements on climate change, gender gap, and aging population for case studies and expert interviews.

\subsection{Procedure}
The case study and interviews were conducted online, utilizing screen-sharing to facilitate the study. At the beginning of each interview, we introduced the motivation behind our work and provided a 15-minute overview of the DataScout workflow. Following this, experts were asked to complete a retrieval task. Specifically,  each expert was given a statement and asked to use DataScout to retrieve relevant data facts. They were free to explore the retrieval space without time constraints and were expected to select 2 to 5 facts that they believed would augment the statement. We encouraged the experts to think aloud and ask questions during the retrieval process. On average, the task took about 30 minutes to complete. After completing the tasks, we conducted semi-structured interviews with the experts to gather feedback on the utility, effectiveness, and usability of DataScout. Overall, each interview lasted approximately 60 minutes. 

In the next, before discussing the experts' comments on our system, we first introduce three example cases given by our experts during the study to show the power of the system.

\begin{figure}[!tbh]
  \centering
  \includegraphics[width=\linewidth]{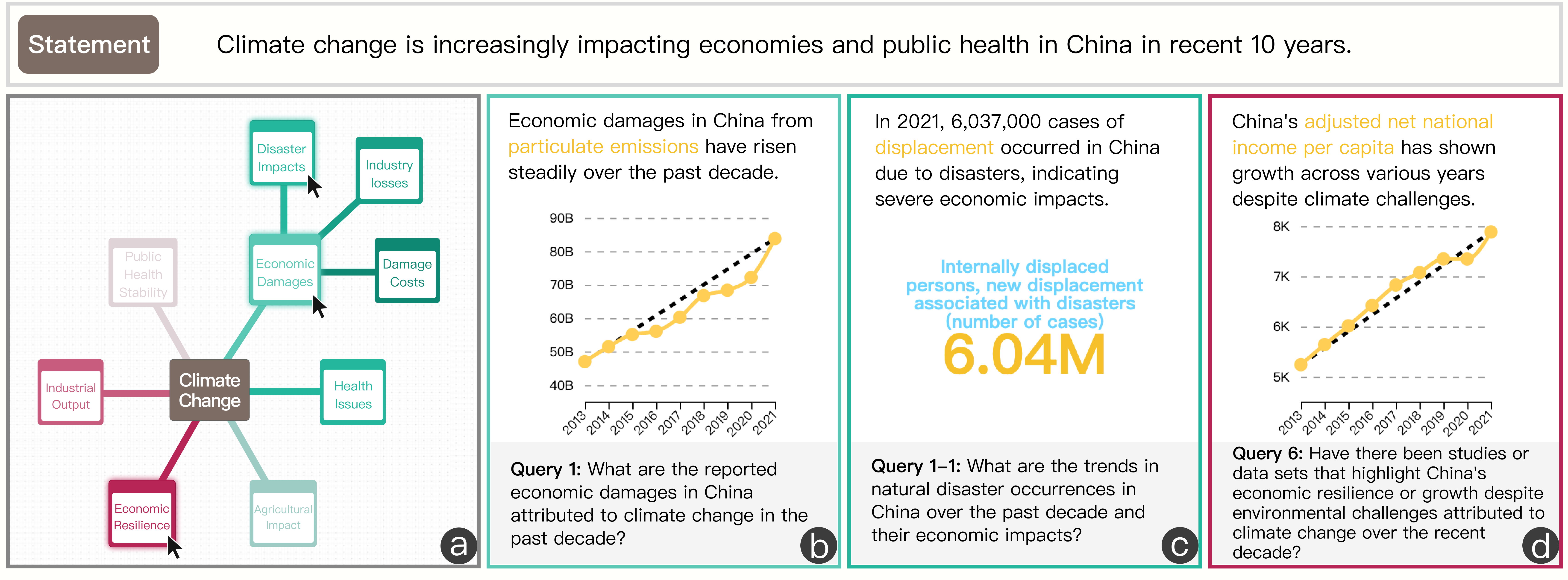}
  \caption{A retrieval space (a) and three retrieved data facts (b, c, d) for the statement \textit{"Climate change is increasingly impacting global economies and public health in China over the past 10 years"}. Data facts (b, c) were retrieved to support the statement. Data fact b shows a growing trend of economic damages from particulate emissions, and data fact c displays the number of displacement cases in China caused by disasters in 2021. Data fact d, retrieved to oppose the statement, shows an upward trend in China's adjusted net national income per capita.} 
  \label{fig:case1}
  \vspace{-1.5em}
\end{figure}

\subsection{Case I: Climate Change}
Fig.~\ref{fig:case1} presents three facts retrieved by E5 for \textit{"Climate change is increasingly impacting global economies and public health in China over the past 10 years"}. First, the query on economic damages attributed to climate change (\textit{Query 1}) retrieved data on particulate emission damage in China from 2013 to 2021. From this, a trend-type fact (\textit{Fact b}) was extracted, revealing that economic damages from particulate emissions in China have steadily risen over the past decade, which strongly supports the statement. \textit{Query 1} was further expanded to explore disaster impact (\textit{Query 1-1}), retrieving data on internally displaced persons and new displacement associated with disasters in China over the past decade. The value-type fact (\textit{Fact c}) shows that the number of cases in the most severe year, 2021, reached 6,037,000. To provide a more balanced argument, a trend-type fact (\textit{Fact d}) was retrieved to oppose the statement, illustrating that despite climate challenges, China's adjusted net national income per capita has shown growth across various years.

\begin{figure}[!tbh]
  \centering
  \includegraphics[width=\linewidth]{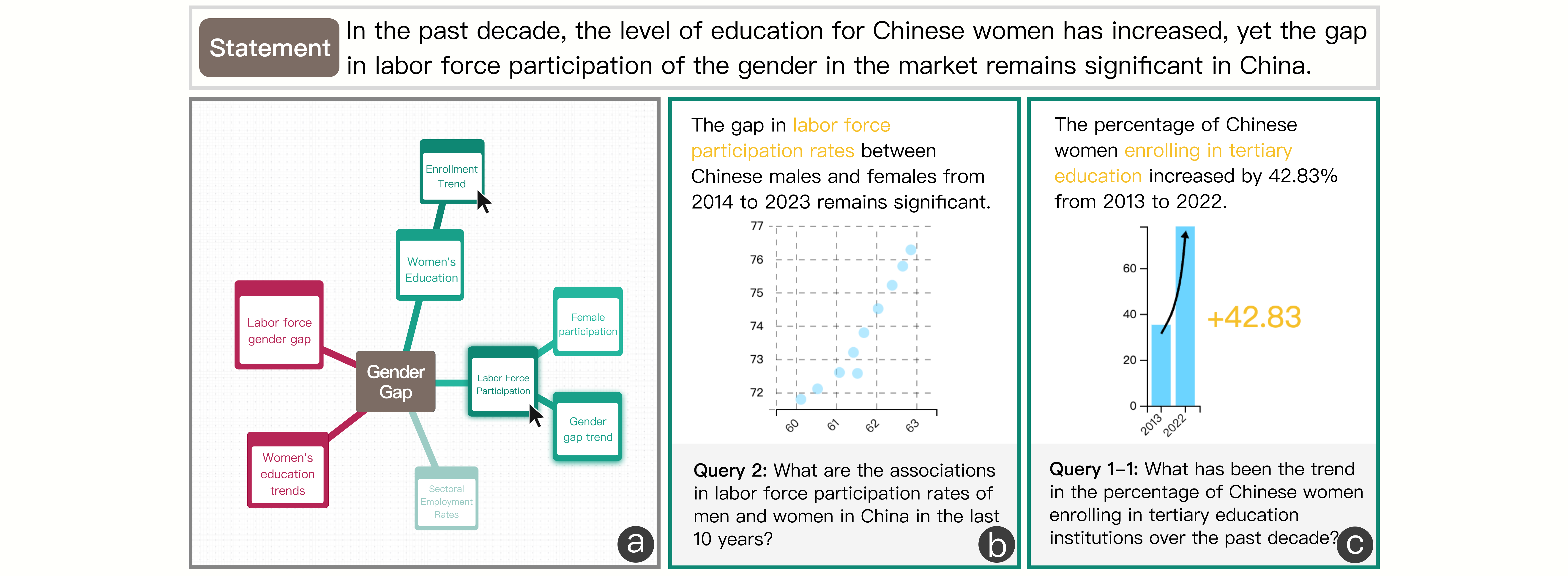}
  \caption{A retrieval space (a) and two retrieved data facts (b, c) for the statement \textit{"In the past decade, the level of education for Chinese women has increased, yet the gap in labor force participation of the gender in the market remains significant in China"}. Data facts (b, c) were retrieved to support the statement. Data fact (b) shows the associations in labor force participation rates of men and women in China, and data fact c shows an increased percentage of Chinese women enrolling in tertiary education from 2013 to 2022.} 
  \label{fig:case2}
  \vspace{-1.5em}
\end{figure}

\subsection{Case II: Gender Gap}
Fig.~\ref{fig:case2} illustrates two facts retrieved by E3 related to the statement: \textit{"In the past decade, the level of education for Chinese women has increased, yet the gender gap in labor force participation in China remains significant"}. 
First, a query on labor force participation (\textit{Query 2}) retrieved an association-type fact, where $measure_1$ is the labor force participation rate for females (x-axis of the scatter plot), and $measure_2$ is the labor force participation rate for males (y-axis of the scatter plot).
This fact indicates that the gender gap in labor force participation between men and women in China remained substantial from 2014 to 2023, strongly supporting the latter part of the statement and showing high relevance. Additionally, \textit{Query 1} was expanded to investigate educational trends (\textit{Query 1-1}), yielding a value-type fact (\textit{Fact c}) showing that the percentage of Chinese women enrolling in tertiary education increased by 42.83\% between 2013 and 2022.

    \begin{figure}[!tbh]
  \centering
  \includegraphics[width=\linewidth]{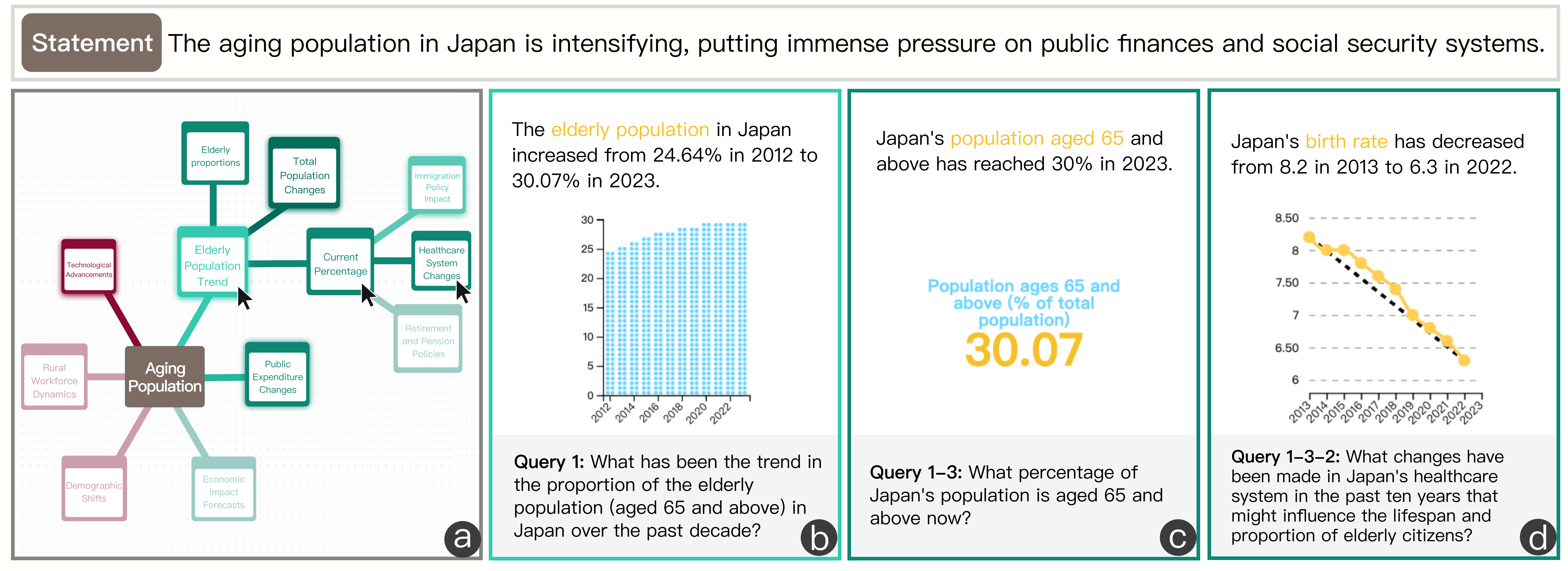}
  \caption{A retrieval space (a) and three retrieved data facts (b, c, d) for the statement \textit{"The aging population in Japan is intensifying, putting immense pressure on public finances and social security systems"}. Data facts (b, c, d) were retrieved to support the statement. Data fact b highlights the increasing proportion of Japan's elderly population (aged 65 and above) over the past decade, while data fact c displays the number of individuals aged 65 in 2023. Data fact d reveals a declining trend in Japan's birth rate.} 
  \label{fig:case3}
  \vspace{-1.5em}
\end{figure}

\subsection{Case III: Aging Population}
Fig.~\ref{fig:case3} presents three facts retrieved by E4 supporting the statement: \textit{"The aging population in Japan is intensifying, putting immense pressure on public finances and social security systems"}. First, a query on the trend of the elderly population of Japan (\textit{Query 1}) provided data on the proportion of individuals aged 65 and above in Japan over the past decade, resulting in a trend-type fact (\textit{Fact b}) that shows the elderly population increased from 24.64\% in 2012 to 30.07\% in 2023. Additionally, \textit{Query 1} was further expanded to determine the current elderly percentage (\textit{Query 1-3}), yielding a value-type fact (\textit{Fact c}) indicating that 30.07\% of Japan's population was aged 65 and above in 2023. 
Further exploration through \textit{Query 1-3} led to a query about healthcare system changes (\textit{Query 1-3-2}). Although the expected answer to the query was not data, the retrieved trend of declining birth rates in Japan from 2013 to 2022 effectively supports the statement.

\subsection{Results}
All experts successfully completed the retrieval tasks. They praised DataScout for its efficiency in retrieving data facts to augment the statements and provided valuable comments, which are summarized as follows:

\underline{\textit{The utility of the stance-based retrieval.}} All experts agreed that the stance-based retrieval feature provides substantial support in the creation of data stories and found the concept highly innovative, e.g., \textit{"I have never seen a tool like this before; it's very novel"} (E5) and \textit{"I really like this idea of retrieving facts to support or oppose a statement"} (E3). When discussing practical scenarios, experts believed that stance-based retrieval is very suitable for validating their current viewpoints. As E4 noted, \textit{"Sometimes I form an initial viewpoint with some supporting data, but I'm unsure whether my perspective is fully rigorous. This feature helps me reflect on both stances of the argument, allowing me to validate the objectivity of my statement and ensure that my argumentation is well-founded."} E3 stated, \textit{"Retrieving supporting or opposing data facts can enrich the narrative of the story."} In summary, stance-based retrieval can help the experts validate the rigor and objectivity of their arguments, leading to more comprehensive stories.

\underline{\textit{The effectiveness of the system.}} All experts noted that the evaluation of fact relevance and stance generally aligned with their own judgments. For example, E4 found a fact in \textit{Query 4} about Japan's high R\&D expenditure, which opposed the claim that the aging population is straining public finances. As E4 noted, \textit{"the system's assessment of low relevance for this fact was appropriate."} Across the retrieval results, all experts selected highly relevant facts with clear stances, confirming the effectiveness of the fact extraction module. We also observed that they all expanded the highlighted nodes. E3 said, \textit{"the highlights help me clarify the direction of my retrieval"}, which demonstrates that the agent’s recommended nodes effectively assisted users in making retrieval decisions. Additionally, the experts praised the idea of expansion for helping them retrieve data facts from multiple perspectives and layers. E3 noted that the system helps creators \textit{"broaden their thinking"}, while E4 pointed out that \textit{"it can improve the efficiency of creating data stories."}

\underline{\textit{The usability of the system.}} The experts found that the interaction is intuitive and easy to understand, particularly appreciating that the mind map layout of the retrieval space clearly displays the reasoning and retrieval process. During the task, we observed that all the experts expanded the nodes with larger sizes and higher color saturation, acknowledging the effectiveness of the color and size encodings for identifying those nodes. However, E3 and E4 felt that \textit{"the distinction in color and size isn't strong enough"} and suggested further optimizing the color contrast and size differences of the nodes to improve recognition efficiency and user experience.

\section{Discussion}
In this section, we discuss the implications derived from our work. Furthermore, we outline several limitations identified during the development and evaluation phases. By highlighting these limitations, we aim to suggest potential directions for future research.

\subsection{Design Implications} 
\noindent\textbf{Applying LLMs to Complex Tasks in Data Storytelling.} To automate data fact retrieval, DataScout draws on the human workflow of retrieving data facts, breaking down the complex retrieval task into specific sub-tasks. Its modular design includes query decomposition, data search, fact extraction, and planning modules, forming an intelligent LLM-based agent. User feedback indicates that DataScout enhances divergent thinking and improves efficiency, showcasing the potential of LLMs in data storytelling. Future research can further extend the application of LLMs to handle more complex tasks, such as automatic writing and infographic generation. By decomposing tasks, LLMs can effectively take on these challenges, streamlining the data storytelling process and boosting efficiency. 

\noindent\textbf{Enhancing Multi-Step Reasoning with Collaborative Interfaces.} In addition to the LLM-based retrieval agent, we designed an interactive retrieval space with a mind-map layout. This interface provides a clear view of the agent's reasoning process and fosters human-AI collaboration that enables users to control the retrieval directions. Future research could explore enhancing this interaction further, enabling users to prompt the LLM for deeper explanations directly within the interface. This collaborative design could also be applied to other domains involving multi-step reasoning, such as Retrieval Augmented Generation (RAG) and text-to-SQL, where an interactive mind-map layout may enhance user engagement and streamline reasoning, thereby improving decision-making efficiency.

\noindent\textbf{Incorporating User's Intent into Intelligence.} In our work, we have demonstrated that stance-based retrieval can effectively help creators validate the rigor of their statements and enrich their arguments with more comprehensive data facts. This highlights the importance of integrating user intent into future intelligent data analysis systems to ensure providing more targeted and context-aware results based on user needs.

\subsection{Limitations and Future Work}
\noindent\textbf{Improving Agent's Accuracy.}
Hallucinations in LLM-powered systems can significantly impact their effectiveness by undermining user trust and reducing reliability. In our study, we identified three issues caused by hallucinations: (1) The caption mismatches the visualization. E4 encountered a chart displaying the complete trend from 1971 to 2023, but the caption only described the trend from 1976 to 1986. While users had the option to edit the caption, E4 noted that such obvious discrepancies could still erode confidence in the system. (2) The fact structure is not standardized. Although we emphasized fact rules in the prompt and implemented fallback mechanisms on the front end, invalid facts still occasionally appeared. (3) The query is non-data-oriented. For example, \textit{Query 1-3-2} in case study III (Fig.\ref{fig:case3}(d)) cannot be directly answered with data. Similar queries also include terms like 'reports' and 'strategies,' which may lead to retrieved data facts with lower relevance to the query. To address these issues, future research could focus on fine-tuning to enhance the LLM's capability to perform specific tasks. Specifically, we can prepare caption-to-facts, stance-to-facts, and query datasets to fine-tune the LLM respectively to ensure that each module of the agent becomes more specialized.

\noindent\textbf{Enriching Data Source.} 
Currently, the data sources are limited to a pre-constructed database. When the statement focuses on a specific topic, the agent generates multifaceted queries, but the range of searchable data fields is restricted. As a result, the retrieved data facts often fail to match the query accurately (as noted by E3-5), and certain data fields are likely to be redundantly described. To address this limitation, a promising future direction is to implement online data search or introduce dynamic database updates. Online search can help users access the most comprehensive and up-to-date data in real-time. However, given the variety of modalities and information sources, future research should focus on enhancing data processing and filtering capabilities. Dynamic database updates refer to identifying a list of official dataset websites in advance and regularly downloading the latest data from them to update the database, which may be a simpler and effective approach.

\noindent\textbf{Supporting Diverse Visualization.}
E3 and E5 pointed out the issue of missing labels on the x-axis and y-axis, which affects the readability of the charts. E4 felt that the charts are too simplistic for presenting complex data and that the system lacks customization options for chart styles. E3 also emphasized the need to display the data source on the charts. In the future, we plan to develop more diverse and flexible chart generation features for captivating data storytelling.

\noindent\textbf{Performing Thorough Evaluation.} We did not compare the user experience of DataScout with other data fact retrieval systems, nor did we evaluate the quality of the retrieved data facts. We would like to perform more in-depth evaluations and assess the system's performance in real-world application scenarios in the future to better understand its practical utility.

\section{Conclusion}
In this paper, we have presented DataScout, an interactive system designed to assist users in retrieving data facts based on stances to enhance statements in data storytelling. The system employs an LLM-based agent to construct a retrieval tree, with its expansion guided by collaboration with the user. The agent decomposes sub-queries based on the user-input statement and stance, then matches these sub-queries to relevant data fields and extracts sub-tables. It applies CoT prompting to extract data facts and assesses their relevance and stance. DataScout provides a retrieval space with a mind map layout that clearly visualizes the reasoning process and retrieval results. 
The proposed system was evaluated via three case studies together with expert interviews. The evaluation results confirmed the overall utility of DataScout and highlighted potential future directions.

\bibliographystyle{ACM-Reference-Format}
\bibliography{reference}




\end{document}



\section{Prompts Used in the Agent}

\begin{center}
    {(1) Prompt for Query Decomposition} 
\end{center}

\begin{framed}
{\ttfamily You are an intelligent assistant tasked with generating sub-queries to retrieve data that \{\{stance\}\} the statement: "\{\{query\}\}". Generate three detailed and relevant sub-queries to help retrieve data facts, evidence, and perspectives. The generated sub-queries should focus on specific metrics or indicators that reflect the stance clearly.\\

For opposing the statement (e.g., "Global environmental conditions are deteriorating"):

- Generate sub-queries that ask about positive trends, improvements, or mitigating factors.

- Example: "Are global CO2 emissions decreasing year by year?"\\

For supporting the statement (e.g., "Global environmental conditions are deteriorating"):

- Generate sub-queries that seek evidence of  negative trends, worsening conditions, or evidence of decline.

- Example: "Is the frequency of extreme weather events increasing worldwide?"\\

Please provide the output in the following format:

\{\{\\
    \doubleindent"queryList": [\\
        \quadindent"Sub-query 1",\\ 
        \quadindent"Sub-query 2",\\
        \quadindent"Sub-query 3"\\
    \doubleindent],\\
    \doubleindent"directionList": [\\
        \quadindent"Overview of Sub-query 1 (1 to 3 words)",\\
        \quadindent"Overview of Sub-query 2 (1 to 3 words)",\\
        \quadindent"Overview of Sub-query 3 (1 to 3 words)"\\
    \doubleindent]
    
\}\}
}
\end{framed}

\vspace{1cm}

\begin{center}
    {(2) Prompt for Text-to-SQL} 
\end{center}

\begin{framed}

{\ttfamily
    The following are the relevant table and its columns for the request:\\

    Table name: \{\{table\_name\}\}
    
    Columns: \{\{relevant\_table[\textquotesingle columns\textquotesingle]\}\}
    
    Columns and their rows: 
    \{\{relevant\_table[\textquotesingle values\textquotesingle]\}\}\\

    Generate a SQL query to retrieve the information for the following request:

    \{\{query\}\}\\

    The SQL query should:
    
    - Be careful to not query for columns that do not exist.
    
    - At least filter out the following series: \{\{relevant\_series\}\}.
    
    - Filter based on the request, especially the mentioned entities.
    
    - Select only existing columns without performing any calculations.
    
    - Limit the result set to 10 rows and 50 columns.
    
    - Output only the SQL query without any additional explanations or text.
}
\end{framed}

\vspace{1cm}

\begin{center}
    {(3) Prompt for Fact Extraction} 
\end{center}

\begin{framed}
{\ttfamily
    You are tasked with extracting data facts from the provided data table that best \{\{stance\}\} the statement.\\
    
    A data fact consists of the following attributes:
    
        \doubleindent- Type: Describes the information type of the fact.
            
        \doubleindent- Subspace: Defines the data scope with a set of filters.
        
        \doubleindent- Breakdown: Temporal or categorical fields for grouping data.
        
         \doubleindent- Measure: Numerical data fields based on which we can retrieve a data value or calculate a derived value, such as sum, average, minimum, or maximum, by aggregating the subspace or each data group.
         
        \doubleindent- Focus: Specific data items in the subspace requiring attention.\\

        You will proceed step by step, making decisions about each attribute in sequence.\\

        Data table: \{\{data\}\}
        
        Statement: \{\{statement\}\}
        
        Query: \{\{query\}\}\\
        
        Step 1: Choose the Type
        
        \doubleindent- Review the data in the tables.
        
        \doubleindent- Identify the most relevant type of data fact that could be used to \{\{stance\}\} the statement. The possible types include:
        
            \quadindent Value: A specific value or measurement.
            
            \quadindent Difference: A comparison between two values.

            \quadindent .......
            
            
            
            
            
            
            
            
        \doubleindent- Provide a brief explanation for your choice of type.\\

        Step 2: Select the Measure
        
        - Based on the chosen type, identify the most relevant numerical data fields and aggregation type.
        
        - Ensure that the measure is directly related to the statement and query.
        
        - Format: [
        
            \doubleindent\{\{\\
                \quadindent"aggregate": "none",  // Options: "none", "max", "min", "sum", "avg"\\
                \quadindent"field": "column name"
                
            \doubleindent\}\},\\
            \doubleindent......
            
        ]
        
        - For Association, provide two measures. For other types, only provide one measure.
        
        - Provide a brief explanation for your choice of measure.\\

        Step 3: Determine the Breakdown
        
        - Consider how the data should be grouped by a temporal or categorical field.
        
        - Choose a breakdown that would make the data fact more compelling in the context of the statement.
        
        - Format: ["column name"]
        
        - For all types, provide one breakdown.
        
        - Provide a brief explanation for your choice of breakdown.\\

        Step 4: Define the Subspace
        
        - Specify the subspace or subset of the data that is most relevant to the statement and query.
        
        - Format: [
                    
                    \doubleindent\{\{
                    
                        \quadindent"field": "column name",
                        
                        \quadindent"value": "xx"
                        
                    \doubleindent\}\},
                    
                    \doubleindent......
                    
                ]
                
        - For Trend, only provide one subspace. For other types, provide one or more subspaces, or leave this list empty.
        
        - For Value, the subspaces should filter out a specific value.
        
        - Provide a brief explanation for your choice of subspace.\\

        Step 5: Focus on the Key Aspect
        
        - Identify the key focus of the data fact.
        
        - The focus value should be a specific 
        value in the selected field.
        
        - Format: [
        
                    \doubleindent\{\{
                    
                        \quadindent"field": "column name"
                        
                        \quadindent"value": "xx"
                        
                    \doubleindent\}\},
                    
                    \doubleindent......
                    
                ]
                
        - For Difference, provide two focus. For Proportion, provide one focus. For other types, provide one focus or return an empty list.
        
        - Explain why this focus is critical to \{\{stance\}\} the statement.\\

        Step 6: Generate a Description
        
        - Generate a concise description (30 words or fewer) that summarizes the data fact and its impact or significance.
        
        - The description should not only describe the fact but also indicate the derived result of the data fact.
        
        - Format: "description"\\

        Final Step: Extract the Data Fact
        
        - After considering all the above attributes, extract the data fact from the tables.
        
        - You should output in this format:
        
            Generated Data Fact: \{\{
            
                \doubleindent"type": "type",
                
                \doubleindent"measure": [...],
                
                \doubleindent"breakdown": [...],
                
                \doubleindent"subspace": [...],
                
                \doubleindent"focus": [...],
                
                \doubleindent"description": "text"
                
            \}\}\\
            
        Repeat the process to extract three data facts.        
}     
\end{framed}

\vspace{1cm}

\begin{center}
    {(4) Prompt for Fact Evaluation} 
\end{center}

\begin{framed}
{\ttfamily
    You are an intelligent assistant tasked with evaluating extracted data facts to determine whether they support or oppose the statement. Think step-by-step and explain the stance (support or oppose) of the data fact towards the statement.\\

        For each data fact provided, please generate probabilities of the data fact supporting and opposing the statement.\\

        \#\# Input:
        
        \doubleindent"Data facts": \{\{facts\}\},
        
        \doubleindent"Statement": \{\{statement\}\}\\

        \#\# Output Format:
        
        [
        
            \doubleindent\{\{
            
                \quadindent"index": 0,
                
                \quadindent"support": [probability],
                
                \quadindent"oppose": [probability],
                
                \quadindent"explanation": [text]
                
            \doubleindent\}\},
            
            \doubleindent......
            
        ]\\

        \#\# Example Output:
        
        [
        
            \doubleindent\{\{
            
                \quadindent"index": 0,
                
                \quadindent"support": 0.76,
                
                \quadindent"oppose": 0.24,
                
                \quadindent"explanation": ...
                
            \doubleindent\}\},
            
            \doubleindent......
            
        ]
}
\end{framed}

\vspace{3cm}

\begin{center}
    {(5) Prompt for Planning} 
\end{center}

\begin{framed}
{\ttfamily
    You are tasked with recommending the next sub-query for data exploration.

        \#\# Goal:
        
        - Given multiple sub-queries and their relevant data facts, recommend the sub-query that should be further explored to retrieve additional relevant data facts that \{\{stance\}\} the given statement: \{\{statement\}\}.
        
        - You should consider the following:
        
        1. Depth of Exploration: Choose the sub-query that could reveal more detailed data facts to \{\{stance\}\} the statement.
        
        2. Stance Alignment: Focus on the sub-query likely to yield data facts that align with the stance.
        
        3. Relevance: Prioritize the sub-query where data facts are most relevant to the statement.
        
        4. Diversity: Consider the sub-query that might offer new or different perspectives on the data.
        
        5. Balance between Known and Unknown: Balance continuing with a sub-query that has already provided useful facts between exploring a potentially unexplored sub-query.\\

        \#\# Observations:
        
        For each sub-query, you have the following details:
        
        1. The index of the sub-query.
        
        2. The sub-query itself.
        
        3. Associated data facts, where each data fact includes:
        
           \doubleindent- The fact itself.
           
           \doubleindent- The stance of the fact towards the statement (whether it supports or opposes the statement).
           
           \doubleindent- The relevance of the fact to the statement.\\

        Here are the sub-queries and their associated data facts:
        
        \{\{queries\_facts\}\}\\

        \#\# Actions:
        
        Based on the observations, recommend one sub-query for further data exploration.\\

        \#\# Output format:
        
        \{\{
        
            \doubleindent"Reasoning": "...",
            
            \doubleindent"Recommend Index": number
            
        \}\}
}   
\end{framed}

\section{Experts Information}
\begin{table}[h!]
\centering
\begin{tabular}{|p{2.8cm}|p{1.9cm}|p{1.9cm}|p{1.9cm}|p{1.9cm}|p{1.9cm}|}
    \hline
    \textbf{Expert ID} & \textbf{E1} & \textbf{E2} & \textbf{E3} & \textbf{E4} & \textbf{E5} \\ \hline
    \textbf{\makecell{Years of experience\\ in data storytelling}}  & 10  & 5  & 5  & 3  & 5  \\ \hline
    \textbf{Role}  & \makecell[l]{researcher,\\ writer}  &  \makecell[l]{journalism\\editor}  & \makecell[l]{data story\\ creator, editor}  & \makecell[l]{data story\\ creator}  & PhD candidate  \\ \hline
    \textbf{Gender}  & Female  & Female  & Female  & Male  & Male  \\ \hline
    \textbf{Major}  & Design  & Journalism  & Journalism  & Journalism  & \makecell{Computer \\Science}  \\ \hline
\end{tabular}
\caption{The background information of all experts involved in the formative study and expert interviews.}
\end{table}